\documentclass[letterpaper,english,aps,prd,twocolumn,superscriptaddress,10pt]{revtex4-1}
\usepackage[T1]{fontenc}
\usepackage[latin9]{inputenc}
\usepackage{amsmath}
\usepackage{amssymb}
\usepackage{wasysym}

\pdfpageheight\paperheight
\pdfpagewidth\paperwidth

\makeatletter

\@ifundefined{textcolor}{}
{
 \definecolor{BLACK}{gray}{0}
 \definecolor{WHITE}{gray}{1}
 \definecolor{RED}{rgb}{1,0,0}
 \definecolor{GREEN}{rgb}{0,1,0}
 \definecolor{BLUE}{rgb}{0,0,1}
 \definecolor{CYAN}{cmyk}{1,0,0,0}
 \definecolor{MAGENTA}{cmyk}{0,1,0,0}
 \definecolor{YELLOW}{cmyk}{0,0,1,0}
}
\makeatother

\usepackage{epsfig}
\usepackage{graphics}  
\usepackage{xcolor}
\definecolor{linkcolor}{RGB}{55,57,154}
\usepackage{hyperref} 
\hypersetup{
	pdfstartview={XYZ},
	colorlinks,
	allcolors=linkcolor,
	bookmarksopen
}
\usepackage{slashed}
\usepackage{cancel} 
\usepackage{babel}

\newcommand{\be}{\begin{equation}}
\newcommand{\ee}{\end{equation}}

\begin{document}

\title{Leptogenesis via the 750 GeV pseudoscalar}

\author{Alexander Kusenko}
\email{kusenko@ucla.edu}
\affiliation{Department of Physics and Astronomy, University of California, Los
Angeles, California 90095-1547, USA}
\affiliation{Kavli Institute for the Physics and Mathematics of the Universe (WPI),
University of Tokyo, Kashiwa, Chiba 277-8568, Japan}

\author{Lauren Pearce}
\email{lpearce@umn.edu}
\affiliation{William I.\ Fine Theoretical Physics Institute, School of Physics
and Astronomy, University of Minnesota, Minneapolis, Minnesota 55455,
USA}
\affiliation{Department of Physics and Astronomy, Valparaiso University, Valparaiso, Indiana 46383}

\author{Louis Yang}
\email{louis.yang@physics.ucla.edu}
\affiliation{Department of Physics and Astronomy, University of California, Los
Angeles, California 90095-1547, USA}

\begin{abstract}
Recently the ATLAS and CMS collaborations have reported evidence of a diphoton excess which may be interpreted as a pseudoscalar boson $S$ with a mass around 750 GeV.  To explain the diphoton excess, such a boson is coupled to the Standard Model gauge fields via $S F \tilde{F}$ operators.  In this work, we consider the implications of this state for early universe cosmology; in particular, the $S$ field can acquire a large vacuum expectation value due to quantum fluctuations during inflation.  During reheating, it then relaxes to its equilibrium value, during which time the same operators introduced to explain the diphoton excess induce a nonzero chemical potential for baryon and lepton number.  Interactions such as those involving right-handed neutrinos allow the system to develop a non-zero lepton number, and therefore, this state may also be responsible for the observed cosmological matter-antimatter asymmetry.
\end{abstract}

\maketitle

\section{Introduction}

Recently the ATLAS and CMS collaborations have reported evidence of a diphoton excess at an invariant mass of $m_{S}\approx750\,\text{GeV}$~\cite{ATLAS1, CMS:2015dxe}.  One possible explanation for the excess is the resonant process $pp\rightarrow S\rightarrow\gamma\gamma$ where $S$ is a new scalar or pseudoscalar field with mass $m_{S}$~\cite{Buttazzo:2015txu,Franceschini:2015kwy,Altmannshofer:2015xfo}.  To produce the signal, this field should couple to the $\mathrm{SU(3)}_{\mathrm{C}}$ field strength tensor $G_{\mu\nu}^{a}$ (for production from gluons) and to the $\mathrm{U(1)}_{\mathrm{QED}}$ field strength tensor $F_{\mu\nu}$ (to enable decays to two photons).

The discovery of beyond-the-Standard-Model physics may have implications for unresolved issues such as the nature of dark matter~\cite{Mambrini:2015wyu,Backovic:2015fnp,Han:2015cty,Dev:2015isx,Han:2015yjk,Park:2015ysf,D'Eramo:2016mgv,Berlin:2016hqw} and the matter-antimatter asymmetry of the universe~\cite{Chao:2016aer,Xiao:2015tja}.  While Ref.~\cite{Chao:2016aer,Xiao:2015tja} considered the impact of a 750 GeV scalar for electroweak scale baryogenesis, we here show that a 750 GeV-scalar pseudoscalar and produce the observed cosmological matter excess during an epoch of relaxation in the early universe, similar to the Higgs- and axion-relaxation leptogenesis scenarios discussed in~\cite{Kusenko:2014lra,Kusenko:2014uta,Pearce:2015nga,Yang:2015ida,Ibe:2015nfa,Gertov:2016uzs,Adshead:2015kza,Adshead:2015jza,Schmitz:2015nqa,Takahashi:2015waa,Takahashi:2015ula,Kainulainen:2016vzv}.

The structure of this paper is as following: In the subsequent section, we introduce a concrete model with a pseudoscalar field with a mass of 750 GeV.  We then discuss how the very operators introduced to explain the LHC diphoton excess can also produce an effective chemical potential in the early universe, if the pseudoscalar field acquire a time-dependent vacuum expectation value (VEV).  Then in section \ref{sec:large_VEV}, we discuss how a large VEV can be produced during the inflationary epoch in the early universe, which will subsequently relax to its equilibrium value.  In section \ref{sec:neutrinos} we discuss the lepton-number violating processes in the early universe which, in the presence of the chemical potential, result in a lepton asymmetry.  The model parameters are restricted by isocurvature constraints and the fact that the entire observable universe is a domain of baryon excess (as opposed to anti-baryonic excess).  These constraints are discussed in section \ref{sec:isocurvature}.  Finally, we present a numerical analysis of the available parameter space in \ref{sec:results}.

\section{The Model and Effective Chemical Potential}
\label{sec:mu_eff}

In order to explain the observed diphoton excess, we supplement the Standard Model (SM) with a real singlet $S$ which interacts via the terms~\cite{Buttazzo:2015txu,Franceschini:2015kwy,Altmannshofer:2015xfo}:
\begin{align}
\mathcal{L}& \supset \tilde{\lambda}_{g}\frac{\alpha_{s}}{12\pi v_{EW}}SG_{\mu\nu}^{a}\tilde{G}_{a}^{\mu\nu}+\tilde{\lambda}_{W}\frac{\alpha}{\pi\sin^{2}\theta_{W}v_{EW}}SW_{\mu\nu}^{a}\tilde{W}_{a}^{\mu\nu} \nonumber \\
&\qquad +\tilde{\lambda}_{B}\frac{\alpha}{\pi\cos^{2}\theta_{W}v_{EW}}SB_{\mu\nu}\tilde{B}^{\mu\nu},\label{eq:SGWB}
\end{align}
where $\theta_{W}$ is the weak mixing angle, and $W$ and $B$ are the $\mathrm{SU(2)}_\mathrm{L}$ and $\mathrm{U(1)}_\mathrm{Y}$ field strength tensors, respectively.  After the Higgs boson acquires a nonzero vacuum expectation value, the Lagrangian contains the couplings
\begin{equation}
\mathcal{L} \supset \tilde{\lambda}_g \frac{\alpha_s}{12 \pi v_{EW}} S G_{\mu \nu}^a \tilde{G}_{a}^{\mu \nu} + \tilde{\lambda}_\gamma \dfrac{\alpha}{\pi v_{EW}} S F_{\mu \nu} \tilde{F}^{\mu \nu},
\end{equation}
where $F^{\mu \nu}$ is the $\mathrm{U(1)}_\mathrm{QED}$ field strength tensor, and $\tilde{\lambda}_\gamma = \tilde{\lambda}_W + \tilde{\lambda}_B$.  Ref.~\cite{Altmannshofer:2015xfo} has explored the parameter space in which this model reproduces the observed excess, finding $\tilde{\lambda}_\gamma = 0.48 \pm 0.08$, although the lower values are in some tension with dijet resonance searches.  Production via gluon interaction is controlled by $\tilde{\lambda}_g \sim 0.1 $ to 1.  
For leptogenesis, we will make use of the operators with the $\mathrm{SU(2)}_\mathrm{L}$ and $\mathrm{U(1)}_\mathrm{Y}$ gauge fields.  These are $\mathcal{O}_5$ operators with an effective scale of $\pi  v_\mathrm{EW} \slash \alpha \tilde{\lambda}_\gamma \sim 10^5 \, \mathrm{GeV}$, although as discussed in Ref.~\cite{Altmannshofer:2015xfo}, these operators are generally constructed from fermions with masses on the TeV scale.

Next, we show that these operators can lead to an effective chemical potential for baryon and lepton number when the $S$ field has a time-dependent vacuum expectation value (VEV).  In the Standard Model, the baryon number and lepton number currents ($j_{B}^{\mu}$ and $j_{L}^{\mu}$) are not conserved; baryon and lepton number can be violated by sphaleron processes.  The divergence of these currents is given by the electroweak anomaly equation
\begin{equation}
\partial_{\mu}j_{B}^{\mu}=\partial_{\mu}j_{L}^{\mu}=\frac{N_{f}}{32\pi^{2}}\left(-g^{2}W_{\mu\nu}^{a}\tilde{W}_{a}^{\mu\nu}+g^{\prime2}B_{\mu\nu}\tilde{B}^{\mu\nu}\right),
\label{eq:anomaly_eq}
\end{equation}
where $N_f = 3$ is the number of families in the Standard Model.

Using the anomaly equation, the terms in \eqref{eq:SGWB} generate a coupling between the pseudoscalar $S$ and the divergence of the $(B+L)$ current,
\begin{equation}
\mathcal{L} \supset -\tilde{\lambda}_{W}\frac{8}{N_{f}v_{EW}}S\partial_{\mu}j_{B+L}^{\mu}=\tilde{\lambda}_{W}\frac{8}{N_{f}v_{EW}}\left(\partial_{\mu}S\right)j_{B+L}^{\mu},
\label{eq:L_with_currents}
\end{equation}
where we have integrated by parts and dropped a total derivative in the second step.  This effective operator is valid when electroweak sphalerons to be in thermal equilibrium; in the early universe, this occurs for temperatures below $T \apprle 10^{12}\,\text{GeV}$ \cite{Daido:2015gqa,Shi:2015zwa}. 

For a patch of the Universe where the VEV $\left< S \right>$ 
is approximately spatially homogeneous but evolves in time, this operator becomes 
\begin{equation}
\mathcal{L}_{O_{5}}=\tilde{\lambda}_{W}\frac{8}{N_{f}v_{EW}}\left(\partial_{0}\left< S \right>\right)j_{B+L}^{0}.\label{eq:chem_ptn}
\end{equation}
which acts as an effective chemical potential
\begin{equation}
\mu_{0}=\tilde{\lambda}_{W}\frac{8}{N_{f}v_{EW}}\left(\partial_{0}\left< S \right>\right)\label{eq:chemical_potential}
\end{equation}
for the $B+L$ charge $j_{B+L}^{0}$.

We assume that the axion-like couplings in \eqref{eq:SGWB}, which are non-renormalizable operators, are effective couplings.  These may be generated by integrating out a loop of fermions which are heavy compared to 750 GeV.  As noted above, we expect these effective operators to break down around the TeV scale.  In the scenario outlined here, we will consider temperatures in the early universe above this.  If this operator is generated by a fermionic loop, these degrees of freedom will typically acquire thermal corrections to their masses, proportional to their coupling times the temperature.  For temperatures $T \gg 10^5 \, \mathrm{GeV}$, the thermal masses will dominate.  Similar finite temperature considerations will apply to other mechanisms of generating the $\mathcal{O}_5$ operators in \eqref{eq:SGWB}.  Therefore, we will use the effective chemical potential
\begin{equation}
\mu_{0} \sim \dfrac{1}{T} \left(\partial_{0}\left< S \right>\right).
\label{eq:chemical_potential2}
\end{equation}

The interpretation of the term in Eq.~\eqref{eq:chem_ptn} as a chemical potential~\cite{Cohen:1987vi,Cohen:1988kt} simplifies the analysis of the asymmetry generation.  However, in some cases, such an interpretation may fail~\cite{Dolgov:1994zq,Dolgov:1996qq}.  The effective chemical potential is a valid approximation when there is a separation of scales: the plasma interactions at temperature $T$ are very rapid on the time scales on which the scalar field is moving.  In this regime, one can introduce two Wilsonian cutoffs, one at some high energy scale $\Lambda_h$ and one at an energy scale $\Lambda_l$, such that $(\partial_0 \left< S \right>/ \left< S \right>) \ll \Lambda_l \ll T$.  One can then integrate out all degrees of freedom outside these two cutoffs and construct an effective theory for the scales between $\Lambda_l$ and $\Lambda_h$.  In this  effective theory, the field $S$ and its time derivative are not propagating degrees of freedom, but slowly varying external parameters.  From the remaining degrees of freedom, describing plasma at temperature $T$, one can construct the Hamiltonian in the usual manner.  The term in Eq.~\eqref{eq:chem_ptn} becomes $\mu_0  n_{B+L}$, where $\mu_0 \propto \partial_0 S$ is the effective chemical potential. 

We observe that this chemical potential, generated by the relaxation of a scalar field, is similar to that in the models considered in Refs.~\cite{Kusenko:2014lra,Kusenko:2014uta,Pearce:2015nga,Yang:2015ida};
however, in this model $\mu_{0}$ depends on the time-derivative of $\left<S \right>$ rather than the time-derivative of $ \left< S ^{2} \right>.$  Consequently, the sign of $\left< S \right>$ is important to determining whether an excess of particle or antiparticles is produced; this will lead to constraints discussed in section \ref{sec:isocurvature} below.

\section{Vacuum Expectation Value during Inflation}
\label{sec:large_VEV}

In the previous section, we demonstrated how the terms between the 750 GeV pseudoscalar field and SM field strengths introduced to explain the observed LHC excess can themselves lead to an effective chemical potential for baryon and lepton number in the early universe, provided that the pseudoscalar field acquired a time-dependent vacuum expectation value.  In this section, we explain how this can naturally occur during inflation.

In addition to the $\mathcal{L}_{O_{5}}$ operator discussed above, the scalar $S$ must have the canonical kinetic term, quadratic coupling, and quartic self-coupling 
\begin{equation}
\mathcal{L}_{S}=\frac{1}{2}\partial_{\mu}S \, \partial^{\mu}S-\frac{1}{2}m_{S}^{2}S^{2}-\frac{1}{4}\lambda_{S}S^{4}\label{eq:L_S}
\end{equation}
for the theory to be renormalizable.  The LHC data suggested that $m_S \approx 750 \, \mathrm{GeV}$. During inflation, the scalar field $S$ can acquire a nonzero vacuum expectation value (VEV), of magnitude $S_{0}\equiv\sqrt{\left\langle S^{2}\right\rangle }$ due to quantum fluctuations. The average initial VEV can be computed through the Hawking-Moss instanton or via a stochastic approach~\cite{Bunch:1978yq,Linde:1982uu,Hawking:1981fz,Starobinsky:1982ee,Vilenkin:1982wt}.  In the massive noninteracting limit ($\lambda_{S}=0$), the average initial VEV has magnitude \cite{Starobinsky:1994bd}
\begin{equation}
S_{0}=\sqrt{\frac{3}{2}}\frac{H_{I}^{2}}{2\pi m_{S}}\approx0.19\frac{H_{I}^{2}}{m_{S}},
\label{eq:VEV_1}
\end{equation}
where $H_{I}\equiv\sqrt{8\pi/3}\Lambda_{I}^{2}/M_{pl}$ is the Hubble parameter during inflation.  For the massless interacting limit ($m_{S}=0$), the VEV is \cite{Starobinsky:1994bd} 
\begin{equation}
S_{0}=\sqrt{\frac{\Gamma\left(\frac{3}{4}\right)}{\Gamma\left(\frac{1}{4}\right)}}\left(\frac{3}{2\pi^{2}\lambda_{S}}\right)^{1/4}H_{I}\approx0.36\frac{H_{I}}{\lambda_{S}^{1/4}}.
\label{eq:VEV_2}
\end{equation}

At the end of inflation, the field rolls down classically to the minimum of its potential. The relaxation of the VEV after inflation provides the time-dependence in the chemical potential (\ref{eq:chemical_potential}).  The evolution of the VEV is governed by the equation of motion,
\begin{equation}
\ddot{S} + 3 H \dot{S} + \Gamma_S \dot{S} + V^\prime(S) = 0,
\end{equation}
where $V(S)$ is the potential from Eq.~\eqref{eq:L_S}, and $\Gamma_S$ is the decay width of the $S$ boson.  The total decay width has been explored in the parameter space for the diphoton excess in Ref.~\cite{Altmannshofer:2015xfo}; it is constrained from below by
\begin{align}
\Gamma(S \rightarrow gg) &= 47 \, \mathrm{MeV} \cdot \tilde{\lambda}_g^2 \left( \dfrac{m_S}{750 \, \mathrm{GeV}} \right)^3,  \nonumber \\
\Gamma(S \rightarrow \gamma \gamma) &= 3.4 \, \mathrm{MeV} \cdot \tilde{\lambda}_\gamma^2 \left( \dfrac{m_S}{750 \, \mathrm{GeV}} \right)^3,
\end{align}
although additional couplings between the $S$ boson and other fields can enhance the decay width.  Ref.~\cite{Altmannshofer:2015xfo} found that in the preferred region of parameters, a decay width between $\mathcal{O}(0.1) -\mathcal{O}(0.01)$ GeV is preferred.
As in the case of the Higgs relaxation, the evolution of the condensate can be treated as classical coherent motion as long as the condensate decay width is not too large~\cite{Enqvist:2014bua,Enqvist:2015sua}.

We note that this potential is invariant under $S \rightarrow -S$, and therefore when a nonzero vacuum expectation value develops, domains with either sign generally occur, separated by domain walls.  In regions where $\left< S \right>$ has different signs, the chemical potential given by \eqref{eq:chemical_potential} also has different signs, which means that whether production particles or antiparticles are biased depends on the sign of the initial vacuum expectation value $\left< S \right>$. 

We note that the potential implied by \eqref{eq:L_S} will also generally acquire finite temperature corrections, generally of the form $\lambda_S S^2 T^2$.  We will focus below on the case in which $\lambda_S$ is small, and therefore these corrections are not significant.

\section{Lepton Number Violating Process - The Standard Seesaw Mass Matrix}
\label{sec:neutrinos}

The results of the previous two sections establish that in the early universe, the pseudoscalar field $S$ naturally acquires a vacuum expectation value and subsequently relaxes to its equilibrium value; furthermore, the very terms introduced to account for the LHC diphoton excess lead to a nonzero chemical potential which can bias the production of particles or antiparticles.  However, this can only occur if the model also includes a lepton-number-violating process.  While there are myriad possibilities for this, we here consider processes mediated by neutrino Majorana mass.  This is motivated by the well-known neutrino seesaw mechanism~\cite{Minkowski:1977sc,Yanagida:1979as,GellMann:1980vs,Mohapatra:1980yp}.

\begin{figure}
\begin{center}
\includegraphics[scale=.5]{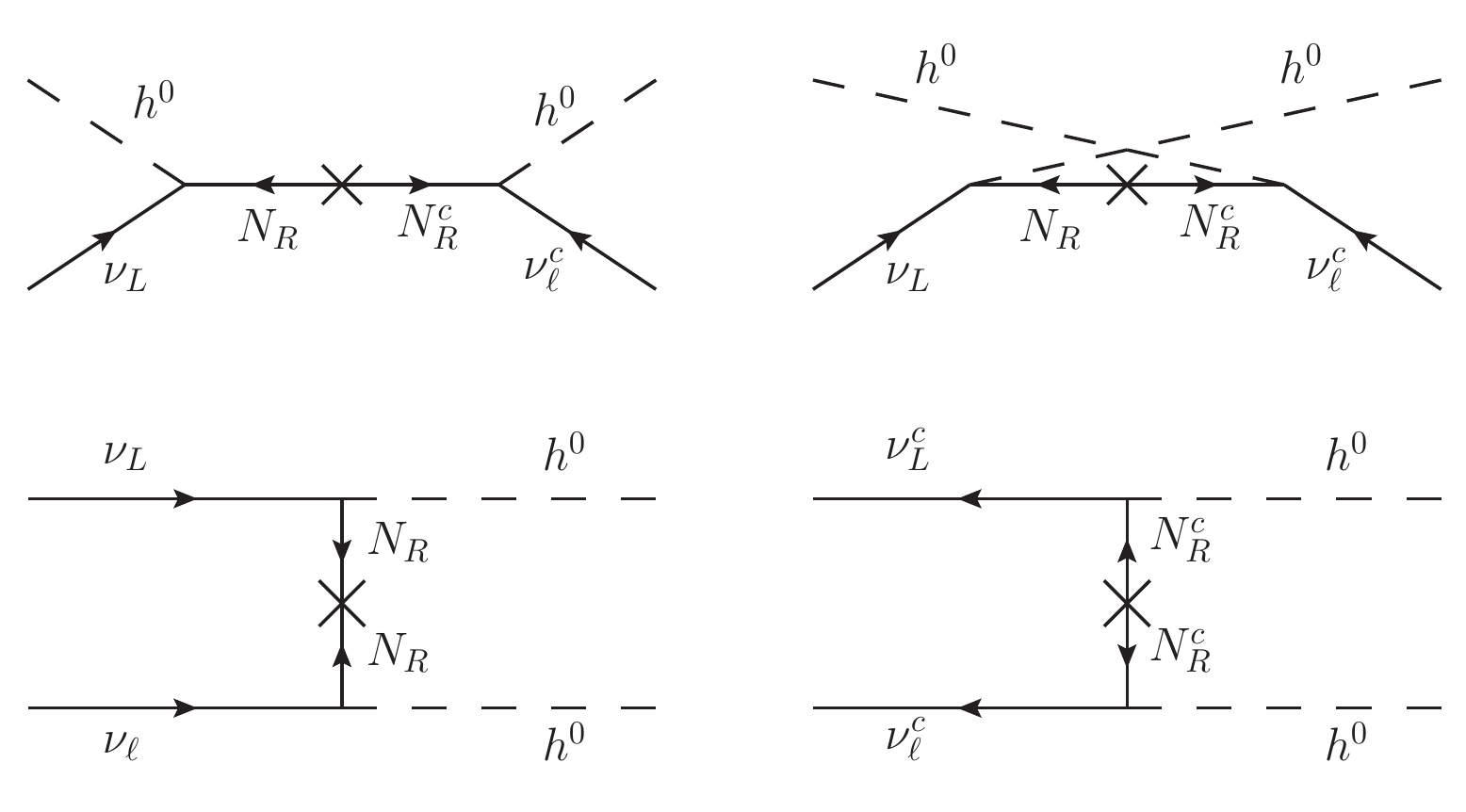}
\end{center}
\caption{Lepton-number violating processes mediated by right-handed neutrinos which generate a lepton asymmetry in the early universe.}
\label{fig:neutrino_diagrams}
\end{figure}

Interactions mediated by a massive right-handed Majorana fermion can violate lepton number, including those shown in Fig.~\ref{fig:neutrino_diagrams}.  These processes are suppressed by the right-handed Majorana mass, which is required to be large in the standard seesaw mechanism.  In the scenario considered here, this is advantageous: if the right-handed Majorana mass is significantly larger than the reheat temperature, then the production of a lepton asymmetry via the production and decay of right-handed neutrinos (as in \cite{Fukugita:1986hr,Covi:1996wh}) is suppressed.  We also note that we ensure that $\mathrm{max}(m_S = \, \mathrm{750} \, \mathrm{GeV},\sqrt{\lambda_S} S_0)$ is smaller than $M_R$, so that decays of the condensate into right-handed neutrinos is also suppressed. 

The thermally-averaged cross section for these processes is~\cite{Yang:2015ida}
\begin{equation}
\sigma_R
\sim m_\nu^2 \slash 16 \pi v_\mathrm{EW}^4 \sim 10^{-31} \, \mathrm{GeV}^{-2},
\label{eq:sigma_R}
\end{equation}
provided that the effective Higgs mass is less than the temperature.  Like the pseudoscalar $S$ field, the Higgs field can acquire a vacuum expectation value during inflation such that $m_\mathrm{H,eff} \sim H_I$, and it will subsequently also relax to its equilibrium position.  This relaxation may occur before or after the relaxation of the pseudoscalar field; however, we will show in section \ref{sec:results} that the asymmetry is generated when the temperature $T \gg H_I$ and so this expression for the cross section is valid.

The heavy Majorana mass does suppress lepton-violating processes in this model as well; however, this is counterbalanced by the chemical potential \eqref{eq:chemical_potential}, which can be large if the pseudoscalar field relaxes to equilibrium rapidly.  We observe that the scattering processes can even be out of thermal equilibrium ($\left< \sigma v \right> n < H$).  The small probability for a single particle to undergo a lepton-number violating interaction is counterbalanced by the fact that we only need to generate a final asymmetry $\mathcal{O}(10^{-10})$ (and the number density remains significant).

In this analysis, we consider these lepton-number violating processes occuring in the plasma of particles produced during reheating.  (We note that because the VEV is trapped at large values until the end of inflation, the relaxation of the pseudoscalar field will generally occur during reheating, without any fine-tuning necessary.)  This is similar to the processes considered in Higgs relaxation analyses such as \cite{Kusenko:2014lra,Yang:2015ida,Gertov:2016uzs} and for axion relaxation in~\cite{Kusenko:2014uta}.  An additional asymmetry is produced by the decay of the condensate itself, as analyzed for Higgs relaxation in  \cite{Pearce:2015nga} and axion relaxation in \cite{Adshead:2015kza,Adshead:2015jza}.

With the inclusion of the right-handed Majorana neutrinos, we have all of the necessary ingredients for relaxation-generated leptogenesis: an effective chemical potential for lepton number, which is generated by a field which acquires a large VEV during inflation that subsequently relaxes to equilibrium, and lepton-number violating processes which can occur during this relaxation.

\section{Domain Size and Baryonic Isocurvature Constraint}
\label{sec:isocurvature}

Equations \eqref{eq:VEV_1} and \eqref{eq:VEV_2} give the magnitude of the average vacuum expectation value of the $S$ field; however, as the vacuum expectation value is produced via quantum fluctuations, different patches of the universe will generally have different VEVs.  In relaxation leptogenesis scenarios, the lepton asymmetry depends on the initial VEV of the field, and therefore, each patch of the Universe could have a different final asymmetry.  As discussed above, not only the magnitude, but also the sign of the final lepton/baryon asymmetry of the Universe is determined by the VEV.  Consequently, in this model the universe would be divided between domains with a matter excess and domains with an antimatter excess. Therefore, the observable universe must fit inside a patch with a single sign of the VEV.  Similar concerns apply to many models of spontaneous baryogenesis~\cite{Cohen:1987vi,Cohen:1988kt}.

We note that $\mathrm{Z}_2$ symmetry of the potential Eq.~\eqref{eq:L_S} is broken by the interactions in Eq.~\eqref{eq:SGWB}.  Through renormalization group equations, these interactions could produce linear or cubic terms in the potential.  Consequently, one of the two vacua could have a lower energy.  If the energy difference is larger than the inflationary Hubble parameter, this vacuum would dominate during inflation, which would suppress the production of domain walls. 

In this section, we will first calculate the constraint from avoiding domain walls in the limit of an exact $\mathrm{Z}_2$ symmetry, which is the most constraining scenario.  Then we will discuss baryonic isocurvature, and show that this leads to stronger constraints than concerns about domain walls.

The characteristic size of one domain can be estimated by the correlation length of the field $S$ during de Sitter expansion. The spatial physical correlation radius $R_{c}$ is given by \cite{Starobinsky:1994bd}
\begin{equation}
R_{c}=H_{I}^{-1}\exp\left(\frac{H_{I}t_{c}}{2}\right),
\end{equation}
where $t_{c}$ is the correlation time. For the massive noninteracting
limit ($\lambda_{S}=0$), 
\begin{equation}
t_{c} = 3 \left(\ln2\right) \frac{H_{I}}{m_{S}^{2}}.
\end{equation}
For the massless interacting limit ($m_{S}=0$), 
\begin{equation}
t_{c}\approx\frac{7.62}{H_{I}\sqrt{\lambda_{S}}}.
\end{equation}

To avoid domain walls, our observable patch of the universe has to be within one domain. This patch has a physical radius
\begin{equation}
R_{RH}\simeq R_{\mathrm{now}}T_{\mathrm{now}}/T_{RH}\sim5\times10^{29}/T_{RH}
\end{equation}
 at the end of reheating. During reheating, the patch grows by a factor
of 
\begin{equation}
R_{RH}/R_{0}\simeq\left(\Lambda_{I}^{4}/T_{RH}^{4}\right)^{1/3}.
\end{equation}
Thus, the patch of our observable universe corresponds to a patch with radius 
\begin{align}
R_{0} & \simeq R_{\mathrm{now}}T_{\mathrm{now}}\left(\frac{T_{RH}}{\Lambda_{I}^{4}}\right)^{1/3} \nonumber \\ &\simeq  R_{\mathrm{now}}T_{\mathrm{now}}\left(\frac{3}{\pi^{3}g_{*}}\right)^{1/12}\frac{M_{pl}^{1/6}\Gamma_{I}^{1/6}}{\Lambda_{I}^{4/3}}\nonumber \\
 & = H_{I}^{-1}R_{\mathrm{now}}T_{\mathrm{now}}\sqrt{\frac{8\pi}{3}}\left(\frac{3}{\pi^{3}g_{*}}\right)^{1/12}\left(\frac{\Lambda_{I}^{4}\Gamma_{I}}{M_{pl}^{5}}\right)^{1/6} \nonumber \\
 & \sim 8\times10^{29}\left(\frac{\Lambda_{I}^{4}\Gamma_{I}}{M_{pl}^{5}}\right)^{1/6}H_{I}^{-1} \nonumber \\
&\approx 2\times10^{23}\left(\frac{\Lambda_{I}}{10^{13}\,\mathrm{GeV}}\right)^{2/3}\left(\frac{\Gamma_{I}}{10^{4}\,\mathrm{GeV}}\right)^{1/6}H_{I}^{-1}
\end{align}
at the end of inflation, where $\Gamma_{I}$ is the inflaton decay rate parameter. For our observable Universe to be within one domain, the correlation radius has to be $R_{c}\gtrsim R_{0}$,
which gives 
\begin{equation}
\frac{H_{I}t_{c}}{2} \gtrsim 53.8 + \frac{2}{3}\ln\left(\Lambda_{I}/10^{13}\,\mathrm{GeV}\right) + \frac{1}{6}\ln\left(\Gamma_{I}/10^{4}\,\mathrm{GeV}\right).
\end{equation}
This imposes constraints 
\begin{equation}
\frac{H_{I}}{m_{S}}\gtrsim 7.19 + 0.05\,\ln\left(\Lambda_{I}/10^{13}\,\mathrm{GeV}\right) + 0.01\,\ln\left(\Gamma_{I}/10^{4}\,\mathrm{GeV}\right)
\end{equation}
for the massive noninteracting case, and 
\begin{multline}
\lambda_{S}\lesssim5.02\times10^{-3}\left[1 - 0.09 \, \ln\left(\Lambda_{I}/10^{13}\,\mathrm{GeV}\right)\right.\\
\left. - 0.02 \, \ln\left(\Gamma_{I}/10^{4}\,\mathrm{GeV}\right)\right]
\end{multline}
for the massless interacting case. 

Furthermore, baryonic isocurvature perturbations are  constrained by the observations of cosmic microwave background \cite{Peebles1987,1987ApJ...315L..73P,Enqvist:1998pf,Enqvist:1999hv,Harigaya:2014tla}.
The upper bound on the baryonic isocurvature perturbation imposed by observations by the Planck satellite is \cite{Ade:2013zuv,Planck:2013jfk,Harigaya:2014tla}
\begin{equation}
\left|\mathcal{S}_{b\gamma}\right|=\left|\frac{\delta Y_{B}}{Y_{B}}\right|\apprle5.0\times10^{-5}.
\end{equation}
To satisfy baryonic isocurvature constraints, and to protect against matter-antimatter domain walls, we require that the observable universe be contained within a single matter domain in which the initial VEV $\left< S \right>$ does not vary significantly.

We will show in Section \ref{sec:results} below that the baryon asymmetry $Y_{B}=n_{B}/s \propto S$ in this particular model where we generate the asymmetry through operator \eqref{eq:chem_ptn}. The isocurvature constraint leads to a condition on the variation of the initial VEV of $S$, 
\begin{equation}
\left|\frac{\delta S}{S_{0}}\right| = \left|\frac{\delta Y_{B}}{Y_{B}}\right| \apprle 5.0 \times 10^{-5}.
\end{equation}
The variation of $\left<S\right>$ about $S_0$ is $\delta S=H_{I}/2\pi$ in a de Sitter space.
Thus, this constraint gives
\begin{equation}
\frac{m_{S}}{H_{I}} \apprle 6.1 \times 10^{-5}
\label{eq:iso_1}
\end{equation}
for the massive noninteracting scenario, and
\begin{equation}
\lambda_{S} \apprle 1.7 \times 10^{-16}
\label{eq:iso_2}
\end{equation}
for the massless interacting scenario.  As expected, these are stronger than the requirement that the observable universe be contained within a domain of the same sign.

We note that in Higgs relaxation scenarios, such as those in Ref.~\cite{Kusenko:2014lra,Yang:2015ida,Pearce:2015nga,Gertov:2016uzs}, it was necessary to introduce new non-renormalizable couplings to evade the baryonic isocurvature constraints.  That is not necessary here, because of the large amount of freedom in the quartic coupling.

\section{Resulting Asymmetry}
\label{sec:results}

In the model introduced, we have shown how a chemical potential is generated, and with the lepton-number violating interactions, the system will approach its equilibrium state of nonzero lepton number.  In general, the system will not reach its equilibrium state during the rapid relaxation of the $S$ field, and so we analyze the generation of the non-zero lepton number with the Boltzmann equation (see the derivation in~\cite{Yang:2015ida}),
\begin{equation}
\dfrac{dn_L}{dt} + 3 H n_L = - \dfrac{2 T^3 \sigma_R}{\pi^2} \left( n_L - \dfrac{2}{\pi^2} \mu_0 T^2 \right),
\end{equation}
where $\sigma_R$ is the thermally-averaged cross section given by \eqref{eq:sigma_R}.

Following the analysis in \cite{Kusenko:2014lra} (setting $M_n = T_\mathrm{rlx}$), we derive an analytic approximation for the resulting asymmetry. During the $S$ field relaxation, we approximate the chemical potential as 
\begin{equation}
\mu_{0} \sim \frac{S_{0}}{T_\mathrm{rlx} t_\mathrm{rlx}},
\end{equation}
using \eqref{eq:chemical_potential2}.  This gives the approximate lepton number density at time $t_{\mathrm{rlx}}$ as
\begin{equation}
n_{L,\,\text{rlx}}\sim\frac{2S_{0}T_{\text{rlx}}}{\pi^{2}t_{\text{rlx}}}\text{min}\left\{ 1,\,\frac{2}{\pi^{2}}\sigma_{R}T_{\text{rlx}}^{3}t_{\text{rlx}}\right\} .
\label{eq:n_rlx}
\end{equation}
The largest asymmetry is produced during the initial relaxation of the $\left< S \right>$ field.  If the oscillations of the $\left< S \right>$ field are not significantly damped, there will be substantial wash-out (as the chemical potential changes sign during the oscillations).  Furthermore, even after the oscillations end and the chemical potential goes to zero, ongoing lepton-number-violating processes will equilibrate the system towards zero lepton number until they go out of equilibrium.  Therefore, we see that a large Majorana mass $M_R$ is again desirable.  During this washout, the system satisfies the approximate Boltzmann equation,
\begin{equation}
\dfrac{dN_L}{dt} = - \dfrac{2 T^3 \sigma_R}{\pi^2}  N_L,
\end{equation}
which leads the following scaling for the lepton number $N_L \equiv n_L a^3$ before and after reheating ends (at $T = T_{RH}$ and $t=t_{RH}\equiv1/\Gamma_{I}$),
\begin{align}
\frac{N_{L}(T)}{N_{L}\left(T_{0}\right)}=\begin{cases}
\exp\left[-\frac{8}{\pi^{2}}\frac{\sigma_{R}T_{RH}^{4}}{\Gamma_{I}}(T^{-1}-T_{0}^{-1})\right]\\
\qquad T\;\mathrm{and}\; T_{0}\ge T_{RH}\\
\exp\left[-\frac{\sqrt{15}}{\pi^{2}}\frac{\sigma_{R}T_{RH}^{2}}{\Gamma_{I}}(T_{0}-T)\right]\\
\qquad T\;\mathrm{and}\; T_{0}\le T_{RH}.
\end{cases}
\end{align}
The asymptotic value of $N_{L}$ at late times is
\begin{equation}
N_{L}(T\rightarrow0)\approx N_{L}(T_{\mathrm{rlx}})\exp\left(-\frac{8+\sqrt{15}}{\pi^{2}}\frac{\sigma_{R}T_{RH}^{3}}{\Gamma_{I}}\right),
\end{equation}
for $t_{\mathrm{rlx}}<t_{RH}$, and 
\begin{equation}
N_{L}(T\rightarrow0)\approx N_{L}(T_{\mathrm{rlx}})\exp\left(-\frac{\sqrt{15}}{\pi^{2}}\frac{\sigma_{R}T_{RH}^{2}T_{\mathrm{rlx}}}{\Gamma_{I}}\right)
\end{equation}
for $t_{\mathrm{rlx}}>t_{RH}$, where $N_{L}(T_{\mathrm{rlx}})$ can
be found using \eqref{eq:n_rlx}. 
From this we find the final ratio of the lepton asymmetry to entropy,
\begin{align}
Y & =\frac{n_{L}}{s}\\
 & \approx\frac{45}{2\pi^{2}g_{*}}\frac{n_{L,\,\mathrm{rlx}}}{T_{RH}^{3}}\frac{N_{L}\left(T\rightarrow0\right)}{N_{L}\left(T_{\mathrm{rlx}}\right)}\left(\frac{a_{\mathrm{rlx}}}{a_{RH}}\right)^{3}\\
 & \approx\frac{45}{2\pi^{2}g_{*}}\left(\frac{2}{\pi^{2}}\right)^{2}\sigma_{R} \frac{S_{0}}{M_{n}} \frac{T_{\mathrm{rlx}}^{5}t_{\mathrm{rlx}}^{2}\Gamma_{I}^{2}}{T_{RH}^{3}}\nonumber \\
 & \qquad\qquad\qquad \times\exp\left(-\frac{8+\sqrt{15}}{\pi^{2}} \frac{\sigma_{R}T_{RH}^{3}}{\Gamma_{I}}\right)\label{eq:estimation_formula}
\end{align}
for $t_{\mathrm{rlx}}<t_{RH}$, and 
\begin{align}
Y & \approx\frac{45}{2\pi^{2}g_{*}}\frac{n_{L,\,\mathrm{rlx}}}{T_{\mathrm{rlx}}^{3}}\frac{N_{L}\left(T\rightarrow0\right)}{N_{L}\left(T_{\mathrm{rlx}}\right)}\nonumber \\
 & \approx\frac{45}{2\pi^{2}g_{*}}\left(\frac{2}{\pi^{2}}\right)^{2}\sigma_{R}\frac{S_{0}}{M_{n}}T_{\mathrm{rlx}}^{2}\exp\left(-\frac{\sqrt{15}}{\pi^{2}}\frac{\sigma_{R}T_{RH}^{2}T_{\mathrm{rlx}}}{\Gamma_{I}}\right)\label{eq:estimation_formula-late}
\end{align}
for $t_{\mathrm{rlx}}>t_{RH}$ at the end of reheating when the oscillation of the scalar field has ended.

In general, the asymmetry is larger for the massive noninteracting case than the massless interacting case, that is, when the $S^2$ term dominates the potential instead of the $S^4$ term.  This may require fine-tuning the quartic coupling to small values, which we discuss below.  For the massive noninteracting case, one can approximate $t_{\mathrm{rlx}} \approx \pi/m_{S}$, provided that $m_{S}\ll H_{I}$ \cite{Starobinsky:1994bd}. $T_\mathrm{rlx}$, the temperature when the field relaxes at time $t_{\mathrm{rlx}}$, is 
\begin{equation}
T_{\mathrm{rlx}}\approx\begin{cases}
T_{RH}\left(\frac{m_{S}}{\pi\Gamma_{I}}\right)^{1/4} & t_{\mathrm{rlx}}<t_{RH}\\
\left(\frac{45}{16\pi^{3}g_{*}}\right)^{1/4}\sqrt{\frac{M_{pl}m_{S}}{\pi}} & t_{\mathrm{rlx}}>t_{RH}.
\end{cases}
\end{equation}
The reheat temperature is $T_{RH}\approx\left(3/\pi^{3}\right)^{1/4}g_{*}^{-1/4}\sqrt{M_{pl}\Gamma_{I}}$ where $M_{pl}$ is the Planck mass.  Using these, the lepton asymmetry can be expressed as
\begin{align}
Y & =\dfrac{n_{L}}{s}\nonumber \\
 & \approx\frac{45}{\sqrt{2\pi^{9}g_{*}}}\sigma_{R}H_{I}^{2}\frac{T_{RH}^{3}}{M_{pl}m_{S}^{2}}\nonumber \\
 & \qquad\times\exp\left(-\frac{8+\sqrt{15}}{\pi^{7/2}}\sqrt{\frac{3}{g_{*}}}\sigma_{R}M_{pl}T_{RH}\right)\\
 & \approx8\times10^{-8}\left(\frac{\sigma_{R}}{10^{-31}\,\mathrm{GeV}^{-2}}\right)\left(\frac{H_{I}}{5\times10^{10}\,\mathrm{GeV}}\right)^{2}\nonumber \\
 & \qquad\times\left(\frac{T_{RH}}{5\times10^{9}\,\mathrm{GeV}}\right)^{3}\left(\frac{750\,\mathrm{GeV}}{m_{S}}\right)^{2}\nonumber \\
 & \;\times\exp\left[-7\times10^{-4}\left(\frac{\sigma_{R}}{10^{-31}\,\mathrm{GeV}^{-2}}\right)\left(\frac{T_{RH}}{5\times10^{9}\,\mathrm{GeV}}\right)\right]
\end{align}
for $t_{\mathrm{rlx}}<t_{RH}$, and 
\begin{align}
 & Y\approx\left(\frac{45}{\pi^{3}g_{*}}\right)^{5/4}\sqrt{\frac{3}{8\pi^{9}}}\sigma_{R}H_{I}^{2}\sqrt{\frac{M_{pl}}{m_{S}}}\nonumber \\
 & \qquad\times\exp\left[-\left(\frac{45}{\pi^{3}g_{*}}\right)^{3/4}\frac{\sigma_{R}}{2\pi^{5/2}}\sqrt{M_{pl}^{3}m_{S}}\right]\\
 & \approx2\times10^{-8}\left(\frac{\sigma_{R}}{10^{-31}\,\mathrm{GeV}^{-2}}\right)\left(\frac{H_{I}}{10^{10}\,\mathrm{GeV}}\right)^{2}\left(\frac{750\,\mathrm{GeV}}{m_{S}}\right)^{1/2}\nonumber \\
 & \;\times\exp\left[-1.3\times10^{-4}\left(\frac{\sigma_{R}}{10^{-31}\,\mathrm{GeV}^{-2}}\right)\left(\frac{m_{S}}{750\,\mathrm{GeV}}\right)^{1/2}\right]
\end{align}
for $t_{\mathrm{rlx}}>t_{RH}$. These estimation formulas agree within one order of magnitude with the numerical results. 

\begin{figure}
\includegraphics[scale=.7]{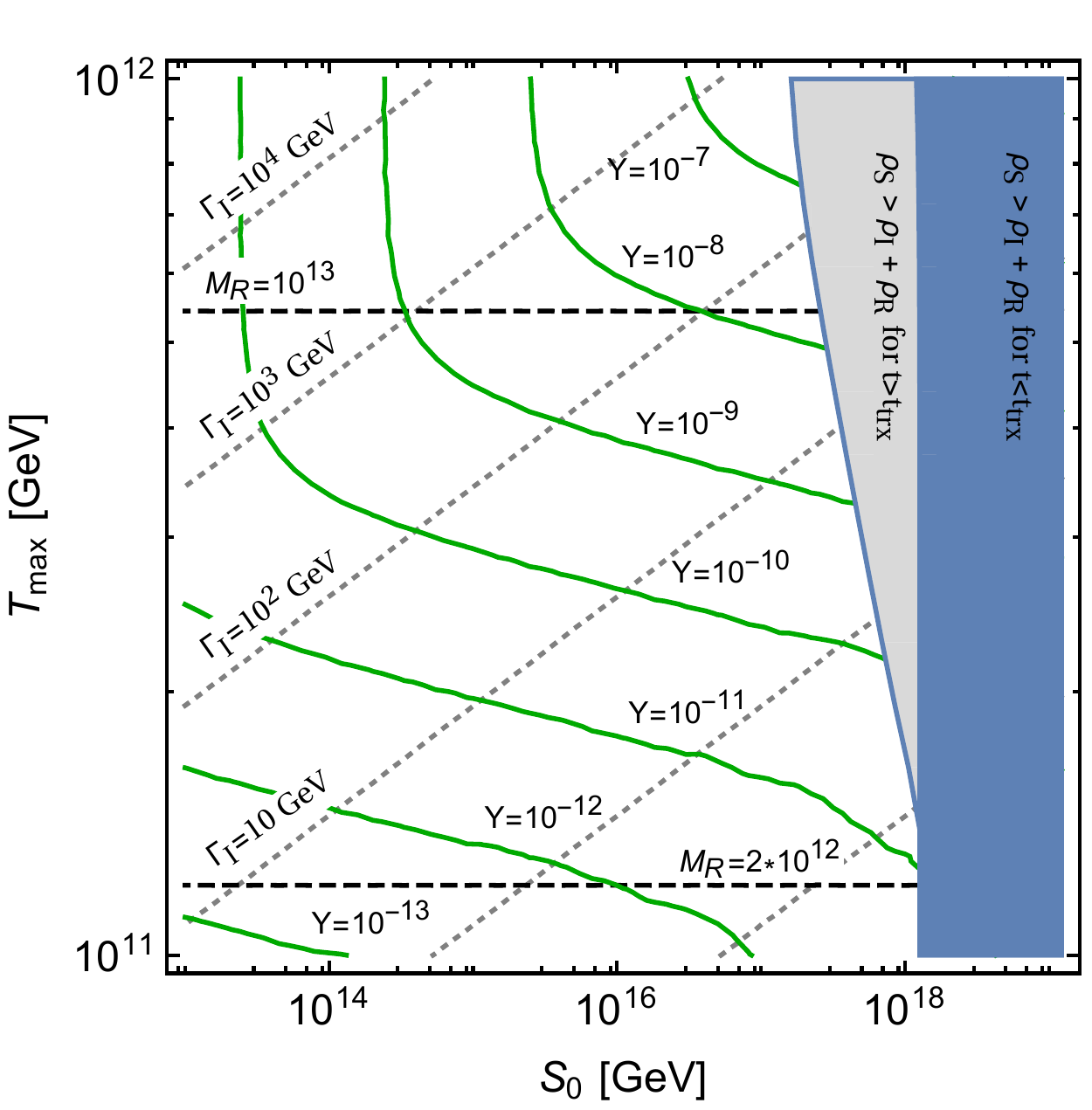}
\caption{The final lepton asymmetry as a function of parameter space for the massive non-interacting scenario.  The dashed lines indicate contours of constant right handed neutrino mass $M_R$ and the dashed lines indicate contours of constant $\Gamma_I$ (the decay rate of the inflaton).  In the shaded region on the right, the pseudoscalar condensate comes to dominate the energy density of the universe.  
}
\label{fig:parameter_space}
\end{figure}

While these are useful analytic approximations, we can also solve the full Boltzmann equation numerically (using the exact cross section $\sigma_R$ given in Ref.~\cite{Yang:2015ida}).  We have done this to explore the available parameter space as a function of $S_0$ and $T_\mathrm{max}$, which is shown in Fig.~\ref{fig:parameter_space}.  We note that the initial VEV of the pseudoscalar field fixes the inflationary scale by equation \eqref{eq:VEV_1} and the relation
\begin{equation}
H_I = \sqrt{\dfrac{8 \pi}{3}} \dfrac{\Lambda_I^2}{M_{pl}}.
\label{eq:HI}
\end{equation}
The inflaton decay parameter $\Gamma_I$ is the fixed by the maximum temperature reached during reheating via~\cite{Weinberg:2008zzc}
 \begin{equation}
T_\mathrm{max} = \left[ \dfrac{30}{\pi^2} \left( \dfrac{3}{8} \right)^{8 \slash 5} \dfrac{2}{3} \sqrt{\dfrac{3}{8 \pi}} \dfrac{\Lambda_I^2 \Gamma_I M_{\mathrm{pl}}}{g_*} \right]^{1 \slash 4},
\end{equation}
where $g_* \approx 107$ is the total number of effectively massless degrees of freedom; we assume that the 750 GeV boson and related fields do not significantly alter this from the Standard Model value.

As noted above, the right-handed neutrinos enable the production of a lepton asymmetry via thermal leptogenesis, with a lepton-to-photon ratio
\begin{equation}
\eta_{th} = \frac{n_{L, th}}{n_\gamma} \approx \epsilon \left( \dfrac{ M_R T_\mathrm{max}}{2 \pi} \right)^{3 \slash 2} \dfrac{e^{-M_R \slash T_\mathrm{max}}}{T_\mathrm{max}^3},
\label{eq:Y_th}
\end{equation}
where we have taken $\epsilon \approx 3 M_R M_\nu \slash 16 \pi v_{EW}^2$ for the CP asymmetry parameter in the lepton sector~\cite{Buchmuller:2005eh}.  (This estimate is found by multiplying the asymmetry parameter by the ratio of non-relativistic right-handed neutrinos to photons at the temperature $T_\mathrm{max}$.)  We note that this is an optimistic estimate for the asymmetry from the thermal decay of right-handed neutrinos, as washout effects and small CP-violating phases can further suppress this.   This will be suppressed by a factor $\sim 30$ due to the entropy production resulting with the Standard Model particles go out of thermal equilibrium.  For the results shown in Fig.~\ref{fig:parameter_space}, we have fixed $M_R$ by setting $\eta_{th} = 10^{-10}$, so that the mechanism discussed here dominates the lepton asymmetry.  (We have verified that for these values, the neutrino coupling constant $y = M_R M_\nu \slash v_{EW}^2$ is in the perturbative regime, taking $M_\nu \approx 0.1 \, \mathrm{eV}$.)  We have shown some contours of $M_R$ on Fig.~\ref{fig:parameter_space}. 

Over the parameter space of interest, the inflationary scale  $\Lambda_I$ ranges from $10^{13} \, \mathrm{GeV}$ (at $S_0 = 10^{14} \, \mathrm{GeV}$) to $10^{14} \, \mathrm{GeV}$ (at $S_0 = 10^{18} \, \mathrm{GeV}$).  Using \eqref{eq:HI}, we see that $T_\mathrm{max} \gg H_I$, which validates the use of the cross section \eqref{eq:sigma_R}.  Contours for $\Gamma_I$, which is significantly smaller, are shown on the plot.  

The blue region on the right of Fig.~\ref{fig:parameter_space} is excluded because the energy density of the pseudoscalar condensate comes to dominate the energy density of the universe.  This can be understood as follows:  Before the pseudoscalar field $S$ relaxes, its energy density is approximately constant and equal to $V_S = m_S^2 S_0^2 \slash 2$.  However, the energy density in the inflaton field and in radiation is decreasing.  The relaxation time, $t_\mathrm{rlx} = \pi \slash m_S$, is less than $1 \slash \Gamma_I$, and therefore the energy density of the inflaton field still dominates the potential, with $\rho_I \approx m_{pl}^2 \slash 6 \pi t^2$.  Imposing $V_S < \rho_I(t_\mathrm{rlx})$ constrains $S_0 < m_{pl} \slash \sqrt{ 3 \pi^3} \approx 10^{18} \, \mathrm{GeV}.$  

We note that constraint \eqref{eq:iso_1} imposes $S_0 \gtrsim 4 \cdot 10^{10} \, \mathrm{GeV}$, which does not eliminate any of the parameter space in which a sufficiently large asymmetry is generated.

In Fig.~\ref{fig:parameter_space}, we have set the decay width of the $S$ boson to 0.1 GeV, near the upper bound of the range suggested by LHC data~\cite{Altmannshofer:2015xfo}.  $\Gamma_I$ is less than 0.1 GeV only beneath the dashed line in the lower right, which means that for much of the parameter space $\Gamma_S < \Gamma_I$, and the $S$ condensate is relatively long lived.  While the condensate is oscillating its energy density is diluted like matter, while after reheating has ended, the plasma loses energy as $a^{-4}$.  Consequently, it is possible for the condensate to come to dominate the energy-momentum density during oscillations, and its decay would significantly re-reheat the universe.  We have excluded this region in gray in the plot (under the approximation that coherent oscillations of the $S$ condensate begin instantly at $t_\mathrm{rlx}$).  

We note that finite temperature corrections may affect $\Gamma_S$.  Decay widths to light Standard Model fermions will be suppressed as $T^{-3}$ due to temperature corrections to the fermionic masses.  However, decay widths to dark sector particles may or may not be similarly affected, depending on the strength of their coupling to electrically charged Standard Model fields.  If the decay width $\Gamma_S$ is further suppressed in the early universe, the pseudoscalar field undergoes coherent oscillations for longer and the bound represented by the gray region becomes more severe.

\begin{figure}
\includegraphics[scale=.6]{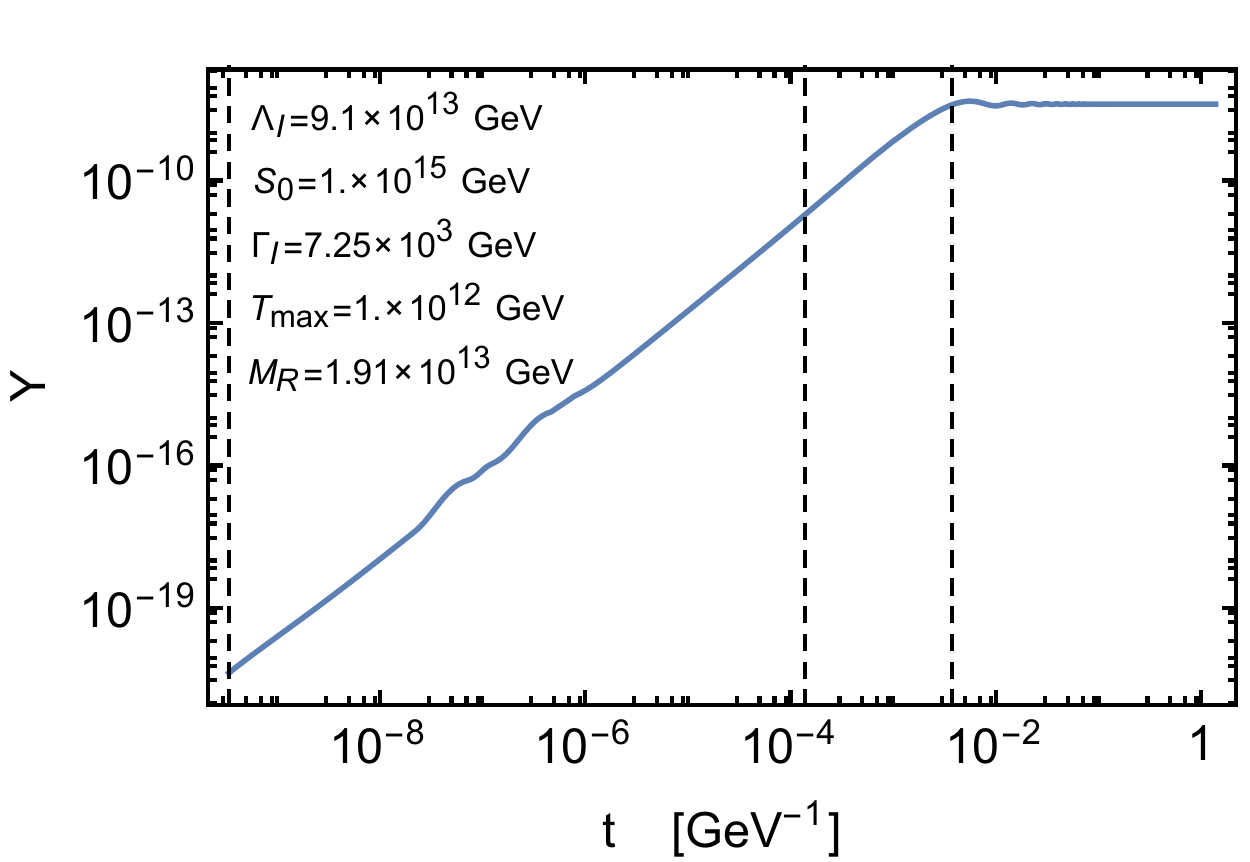}
\caption{The evolution of the lepton asymmetry as a function of time for the parameters indicated.  The dashed vertical lines indicate the time of maximum temperature, the beginning of the
radiation dominated era, and the first time the $S$ VEV crosses zero, from left to right.}
\label{fig:Y_evol}
\end{figure}

In Fig.~\ref{fig:parameter_space}, we have restricted our maximum temperature to $10^{12} \, \mathrm{GeV}$.  As noted above, at this temperature electroweak sphalerons go out of thermal equilibrium, and therefore, the replacement of the couplings in Eq.~\eqref{eq:SGWB} with the baryon and lepton current in \eqref{eq:L_with_currents} is invalid.  However, as shown in Fig.~\ref{fig:Y_evol}, most of the lepton asymmetry is produced after the temperature has reached its maximum value.  Therefore, there should be some parameter space available at larger temperatures.  However, a complete analysis of the situation in which the electroweak sphalerons are in thermal equilibrium during only part of the evolution of the pseudoscalar condensate is beyond the scope of this paper.

\begin{figure}
\includegraphics[scale=.6]{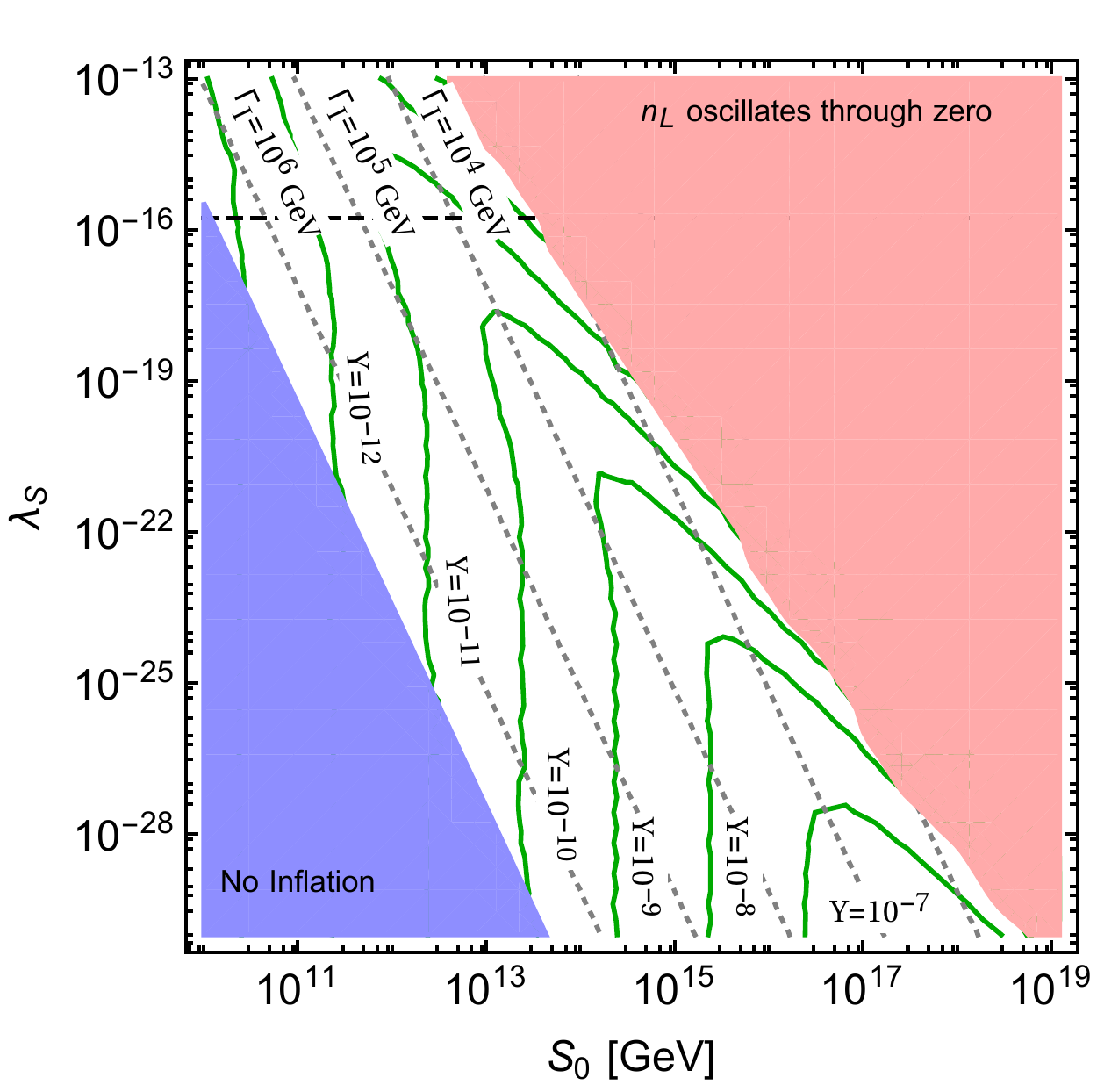}
\caption{The final lepton asymmetry as a function of parameter space for the quartic potential and $T_\mathrm{max} = 10^{12} \, \mathrm{GeV}$.  In the lower left section, $\Gamma_I > H_I$, and so there would be no inflationary epoch.  In the upper right section, $n_L$ oscillates through zero.  The black dashed line shows the constraint \eqref{eq:iso_2}, while the gray dotted lines show contours of constant $\Gamma_I$.}
\label{fig:quartic_param_space}
\end{figure}

As we mentioned above, this analysis is for the massive non-interacting scenario, in which the potential is dominated by the quadratic term.  This is valid for quartic terms $\lambda_S < 2 m_S^2 \slash S_0^2$.  For this hold to $S_0 = 10^{18} \, \mathrm{GeV}$ requires $\lambda_S < 10^{-30}$, or to hold to $S_0 = 10^{15} \, \mathrm{GeV}$ requires $\lambda_S < 10^{-24}$.  It would be difficult to arrange such small couplings without some degree of fine-tuning, unless the $S$ boson is embedded in a larger scalar sector, such that the potential has a flat direction.  (Note that in contrast to Affleck-Dine baryogenesis~\cite{Affleck:1984fy,Dine:1995kz}, such a flat direction would not need to carry lepton or baryon number.)  

In the massless interacting scenario, we have an additional degree of freedom corresponding to $\lambda_S$.  To explore the parameter space, we set $T_\mathrm{max} = 10^{12} \, \mathrm{GeV}$, since in the massive case a significant asymmetry was produced only near this limit.  Fixing $\eta_{th} = 10^{-10}$ sets $M_R \approx 2 \cdot 10^{13} \, \mathrm{GeV}$, and as before, we have taken $\Gamma_S = 0.1 \, \mathrm{GeV}$.

The results are shown in Fig.~\ref{fig:quartic_param_space}.  The scale of inflation is set by
\begin{equation}
\Lambda_I^2 = M_{pl} S_0 \sqrt{\dfrac{3	\Gamma(1 \slash 4)} {8\pi \Gamma(3 \slash 4)}} \left( \dfrac{2 \pi^2\lambda_S}{3} \right)^{1 \slash 4},
\end{equation}
In this plot, it ranges from $\mathcal{O}(10^{12} \, \mathrm{GeV})$ near $S_0 = 10^{10} \, \mathrm{GeV}$ and $\lambda_S = 10^{-13}$ to $\mathcal{O}(10^{15} \, \mathrm{GeV})$ near $S_0 = 10^{18} \, \mathrm{GeV}$ and $\lambda_S = 10^{-30}$.  Contours of constant $\Gamma_I$ are shown; we have marked the region in which no inflation occurs because $\Gamma_I > H_I$ in the lower left.

In the upper right, washout is strong enough that the lepton asymmetry $n_L$ oscillates through zero.  We note that the quartic potential is steeper than the quadratic potential; consequently, the pseudoscalar VEV relaxes to its equilibrium value faster.  This increases $\mu_0$ given by Eq.~\eqref{eq:chemical_potential2}, but the system has less time in which to generate the asymmetry.  Furthermore, the VEV continues to evolve quickly during its oscillation, leading to relatively large chemical potentials during this epoch.  Therefore, washout is a more severe problem in the quartic potential.  This can be alleviated if the pseudoscalar field $S$ were to acquire a larger decay width $\Gamma_S$; this increases the effective friction which decreases the amplitude of the oscillation (in addition to slowing the relaxation).  
 
We see that generating a sufficiently large asymmetry generally requires a small $\lambda_S$, although not quite a small as required for the quadratic term to dominate the potential.  A sufficiently large asymmetry can be generated with $\lambda_S \sim 10^{-20}$ if $S_0 \sim 10^{15}$.  This is more stringent than limit \eqref{eq:iso_2}, which is shown by the black dashed line.

As mentioned above, a larger asymmetry can be generated if we consider higher temperatures.  We note, however, that equation \eqref{eq:Y_th} would imply a non-perturbative coupling for the neutrino sector for $T_\mathrm{max} > 1.4 \cdot 10^{13} \, \mathrm{GeV}$.  However, a small CP-violating phase in the neutrino can relax this.

\section{Conclusions}

Observations of the 750 GeV diphoton excess at the LHC has motivated the consideration of a pseudoscalar field which couples to the electromagnetic field strength~\cite{Buttazzo:2015txu,Franceschini:2015kwy,Altmannshofer:2015xfo}.  To generate this operator, the pseudoscalar field $S$ should couple to the fundamental $\mathrm{SU(2)}_\mathrm{L}$ and/or $\mathrm{U(1)}_\mathrm{Y}$ field strengths.  The first of these couplings generates a chemical potential for lepton and baryon number in the early universe, as the pseudoscalar field relaxes from a large vacuum expectation value naturally generated by quantum fluctuations during inflation.  In the presence of lepton-number violating interactions, such as those mediated by heavy right-handed neutrinos, a nonzero lepton asymmetry is produced, which is transferred to baryons via electroweak sphalerons.  

We have explored the parameter space in which a sufficiently large asymmetry is generated via this mechanism for both a quadratic and quartic potential.  In particular, there are regions of parameter space in which a sufficiently large asymmetry is generated through this mechanism while the asymmetry generated by thermal leptogenesis is insufficient.  We have also considered constraints from the condition that the entire observable universe have a particle excess (in contrast to an antiparticle excess) and baryonic isocurvature observations.  These do not restrict the available parameter space.  This is in contrast to the Higgs relaxation models~\cite{Kusenko:2014lra,Yang:2015ida,Gertov:2016uzs}, in which additional non-renormalizable couplings were required to satisfy isocurvature constraints.

A.K.~and L.Y.~were supported by the U.~S.~Department of Energy Grant
DE-SC0009937.  A.K.~was also supported by the World Premier International Research Center Initiative (WPI), MEXT, Japan.  L.~P.~was partially supported from the DOE grant DE-SC0011842 at the University of Minnesota.

\bibliographystyle{apsrev4-1}
\bibliography{Reference}

\end{document}